# Direct determination of 2D Momentum Space from 2D Spatial Coherence of Light using a Modified Michelson Interferometer

S. V. U. Vedhanth and Shouvik Datta [a]

*Department of Physics, Indian Institute of Science Education and Research, Pune 411008, Maharashtra, India*

Authors to whom correspondence should be addressed:

[a] shouvik@iiserpune.ac.in

Momentum space distributions of photons coming out of any light emitting materials/devices provide critical information about its underlying physical origin. Conventional methods of determining such properties impose specific instrumentational difficulties for probing samples kept within a low temperature cryostat. There were past studies to measure one dimensional (1D) coherence function which could then be used for extracting momentum space information as well as reports of measurements of just two dimensional (2D) coherence function. However, all of those are associated with additional experimental complexities. So, here we propose a simpler, modified Michelson interferometer based optical setup kept at room temperature outside the cryostat to initially measure the 2D coherence function of emitted light, which can then be used to directly estimate the 2D in-plane momentum space distribution by calculating its fast Fourier transform. We will also discuss how this experimental method can overcome instrumentational difficulties encountered in past studies.

**I. INTRODUCTION**

Physical mechanisms behind any light emission process are often marked with increased spatio-temporal coherence in the system, like lasing[1–8] and formation of Bose-Einstein Condensate (BEC)[9–16] of 'bright' excitons, polaritons, photons. These excitons are formed when electrons and holes are excited within a semiconductor. These oppositely charged particles are mutually attracted by Coulomb interaction to form bound electron-hole pairs as hydrogen like artificial atoms within these solids. If not spin forbidden or 'dark', some of these excitons can radiatively recombine and generate light. In case of lasing, stimulated emission can also lead to enhanced emission of light which can be temporally coherent. However, during the formation of exciton or polariton BEC, there will also be a spontaneous, long-range spatial order when all particles in the system attain quantum degeneracy below the critical temperature. These processes can be probed by measuring $1^{st}$ order spatial coherence function using optical interferometry. Various reports on probing such events have already been made using modifications of Young's double slit experiment[17–20], Michelson Interferometer[9,21–24] and Mach-Zehnder Interferometer[25–27]. There were also attempts to measure both spatial and temporal coherence using the same setup[24]. Another way to probe these processes, especially the formation of BEC, is to monitor the momentum space or the k-space of the system to see the onset of spatially long range order/correlation. When a BEC is formed, it will be marked by an increase in long range spatial order and consequently by a characteristic narrowing in momentum space. Such 2D momentum space measurements are often done by scanning the Back Focal Plane (BFP) image of a lens[28–30]. Now, for most systems like excitonic BEC, the samples are generally needed to be kept in a cryostat to go below the critical temperature to observe a BEC. In such cases, performing Young's double slit experiment with varying slit widths and adjustable slit separations kept within the cryostat to measure 1st order spatial correlation to estimate the coherence function can be somewhat challenging in terms of instrumentation. BFP imaging also require placing a high numerical aperture lens very close to the sample. All of these actually increase experimental complications for low temperature samples placed inside a cryostat. These include additional heat loads on the sample chamber and/or to the sample itself due to its proximity with various optical elements, some of which are at ambient temperature. Therefore, one way to measure these optical coherence functions is to extract the light emitted from the sample outside the cryostat and then perform the experiment[24] with a modified optical interferometry setup kept in ambient temperature and pressure and at a distance away from the sample.

In general, optical coherence [31–33] is the property that describes the phase stability of any light, either in the temporal or spatial domain. The temporal (spatial) coherence is characterized by the coherence time $\tau_c$ (with coherence length $L_c$ in the direction transverse to the propagation of light). The coherence time (length) is the duration (transverse distance) over which the phase of the wave remains stable. Let's say that we know the wave's phase at $t_1$ ($r_1$), then the phase at the same position (time) but at different time $t_2$ (position $r_2$) will be known with a high degree of certainty when $|t_2 - t_1| \leq \tau_c$ ($|r_2 - r_1| \leq L_c$), and with a very low degree when $|t_2 - t_1| \geq \tau_c$ ($|r_2 - r_1| \geq L_c$). Ordinarily, optical coherence can be quantified by the First Order Coherence Function, $g^{(1)}(r_1, t_1, r_2, t_2)$ defined as,

$$g^{(1)}(r_1, t_1, r_2, t_2) \equiv g^{(1)}(s, \tau) = \frac{\langle \varepsilon^*(r_1,t_1)\varepsilon(r_2,t_2)\rangle}{\sqrt{\langle|\varepsilon(r_1,t_1)|^2\rangle\langle|\varepsilon(r_2,t_2)|^2\rangle}} \quad (1)$$

where $s = \delta r = (r_2 - r_1)$ is the spatial separation in the transverse direction and $\tau = \delta t = (t_2 - t_1)$ is the temporal separation along the propagation direction. This (temporal) spatial coherence is calculated from this equation at a particular (position) time. The modulus of the first order coherence function, $|g^{(1)}(s,\tau)|$ is called the degree of first-order correlation or coherence [31-33]. This is 1 for a completely coherent, ideal light source, 0 for a completely incoherent light source and varying in between for any partially coherent source. For any interference of two light waves, the visibility of the fringe pattern can be related to the degree of coherence as

$$\text{Visibility} = \frac{I_{max}-I_{min}}{I_{max}+I_{min}} = \frac{2(\langle I_1\rangle\langle I_2\rangle)^{1/2}}{\langle I_1\rangle+\langle I_2\rangle}|g^{(1)}(s,\tau)| \quad (2)$$

For the interfering sources of equal average intensities, one gets

$$\text{Visibility} = |g^{(1)}(s,\tau)| \quad (3)$$

Interestingly, the coherence functions as quantitative measures of statistical fluctuation of 'phase' are associated with the Wiener-Khintchine Theorem[33] of statistical physics. It states that the energy spectrum of the source is the Fourier Transform of the first order temporal coherence as

$$F(\omega) = \frac{1}{2\pi}\int_{-\infty}^{\infty} g^{(1)}(\tau)e^{i\omega\tau}d\tau \quad (4)$$

Similarly, obtaining the first order spatial coherence function as a function of spatial separation can also help us in probing the momentum space distribution.

$$F(k) = \frac{1}{2\pi}\int_{-\infty}^{\infty} g^{(1)}(s)e^{iks}d(s) \quad (5)$$

There were past experiments by L.V. Butov et al.[27] to measure the spatial coherence function and thus calculating the k-space, but it was executed along only one spatial dimension using a Mach-Zhender interferometer. There were also past studies[9,34] to measure 2D spatial coherence function using interferometry as well. One of these employed Cat-Eye retroreflector (CER) and the other used a dove prism. However, these curved retroreflectors can give substantial spherical aberrations, specifically in case of spherical CER[35,36] or polarization dependent phases in case of Corner-Cube CER[37] which uses oblique configuration of mirrors. However, in our experiments, we would like to have a control over the phase differences. So we wanted remove any such unintentional optical phases introduced by the measurement set ups. In case of dove prism there can also be manufacturing imperfections leading to lateral displacement of the output beam[38]. Due to these problems with the CER and the dove prism, an extra care must be taken for their optical alignments. Moreover, these optical elements are also costlier. Nevertheless, the full gamut of the relationship between 2D coherence function and 2D momentum space distribution of light has rarely been studied.

Anyway, to determine this 2D momentum space with the help of Weiner-Khintchine theorem, we needed to measure the 2D spatial coherence function over a certain cross-sectional area of the light beam. So we first tried to measure and generate a 2D map of the coherence function using the modified Michelson Interferometry[24]. We then calculate the 2D momentum space for a laser diode as a function of bias current where the coherence function of the emitted light can be modified from LED mode, when the light will be incoherent and then in the lasing mode, when they will be coherent, by changing the bias current over the threshold value for lasing.

## II. EXPERIMENTAL SETUP

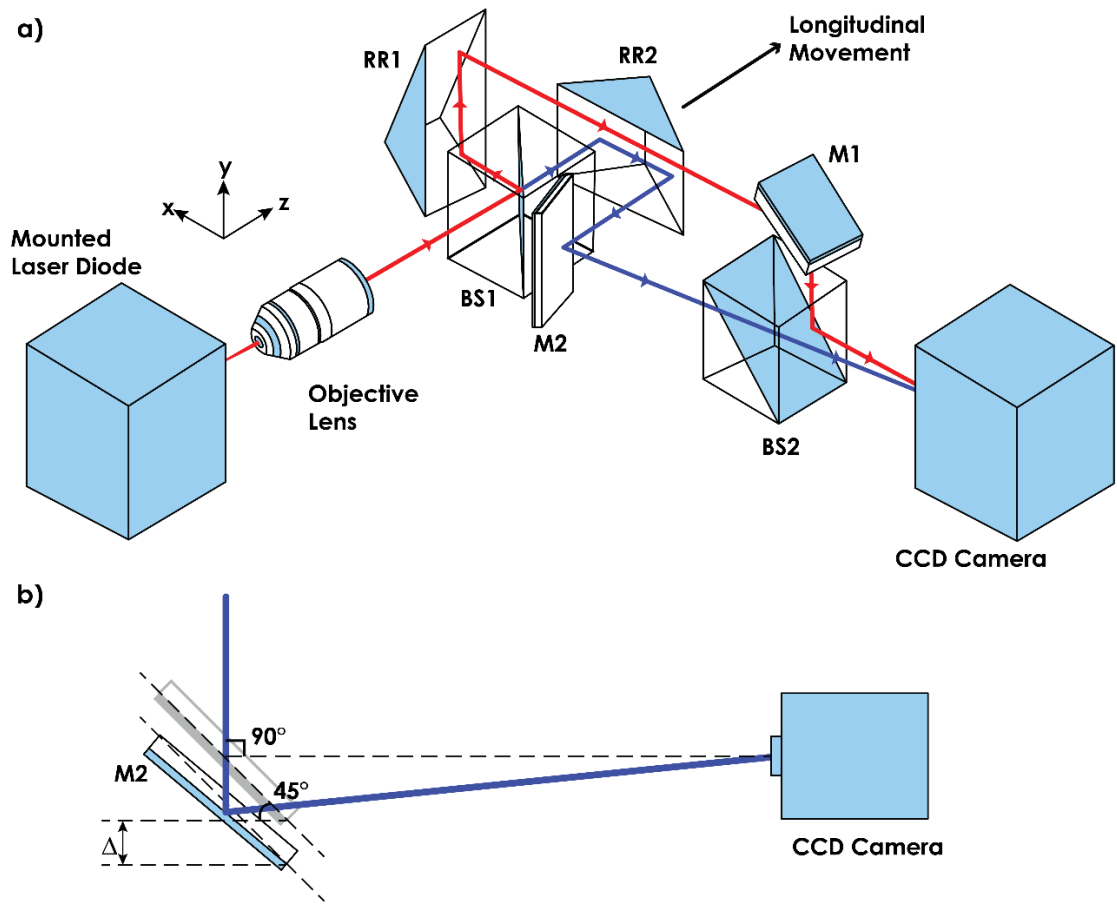

**FIG. 1.** The above image is a schematic representation of our 2D spatial coherence measurement setup. The directions are marked with respect to the direction of propagation of light. The lower image shows how the tilt is introduced into the beam. Gray coloured mirror represents the condition of no tilt.

The schematic representation of our setup is shown in Fig.1. This setup is a modified Michelson interferometer[24] where the usual mirrors in the setup are replaced with triangular retroreflectors (RR1 and RR2) with planar surfaces. The 50:50 beam splitter BS1 is CCM1-BS013/M from Thorlabs. The mirrors M1, M2 are silver protected mirrors PF10-03-P01 from Thorlabs, and the silver-coated retroreflectors are procured from Holmarc, India.

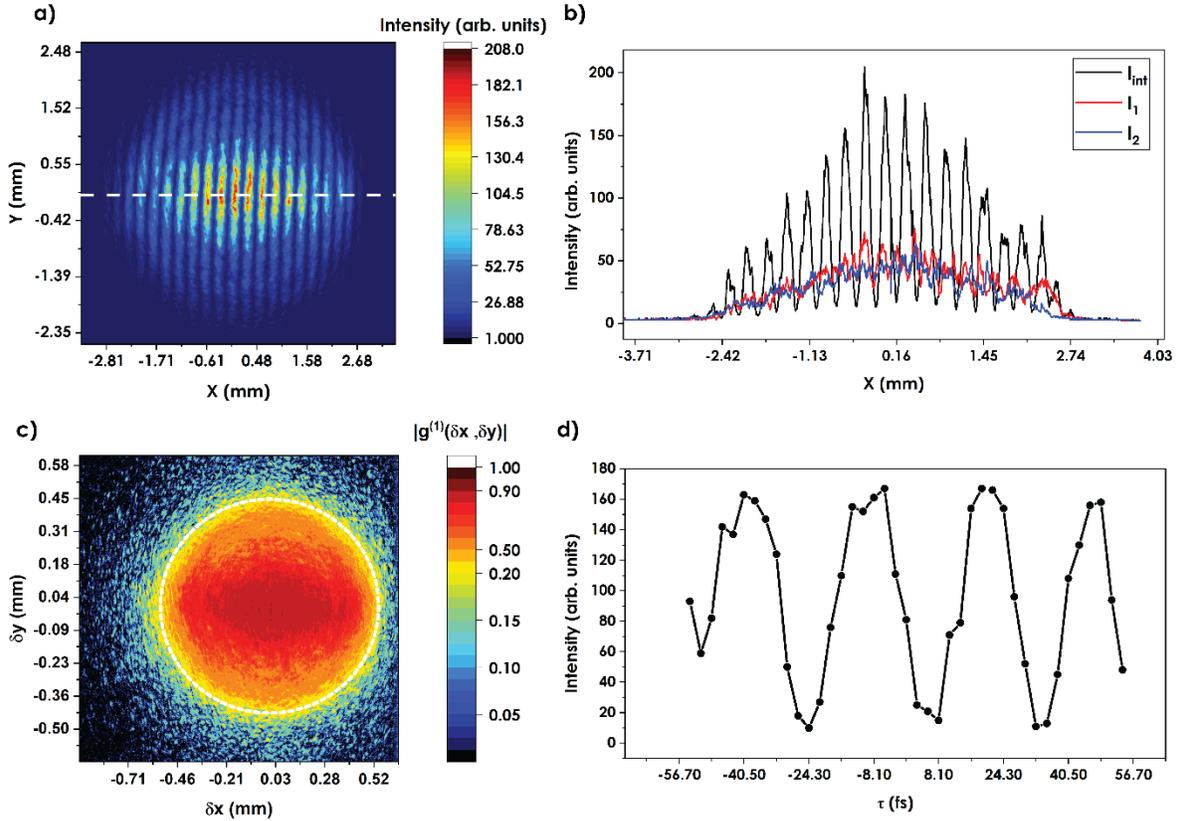

**FIG. 2.** (a) This shows one of the interference patterns observed for 30 mA and (b) shows its profile along x axis. (c) is the 2D coherence function obtained in case of 30 mA. The dotted circle shows the region selected for the k-space calculation. (d) shows the I($\tau$) obtained for a pixel near to ($\delta x=0, \delta y = 0$) at the same bias current.

In this case, the plane of reflections from the retroreflectors are not coplanar as earlier[24] and reflected beams from RR1 and RR2 are at different heights. As a result, we needed these two additional mirrors M1 and M2. In general, these retroreflectors can prevent any back reflection from the source. Use of two such retroreflectors can also provide the optical setup with a certain tilt resistance[39] capabilities. The light source used is HL6358MG (Thorlabs), an AlGaInP based multi-quantum well commercial single-mode laser diode with the peak wavelength at 645 nm and the threshold for lasing at 25 mA. It is mounted on TCLDM9 mount, which helps to control the diode's operating temperature. It is ran using the ITC4001 controller from Thorlabs. The retroreflector RR2 is attached to an xyz-stage through a piezo actuator so that it can be moved longitudinally to introduce temporal phase difference. The light from the diode is collimated using an objective lens of magnification of 10x. The interference is detected using a simple CCD camera based BC106NVIS/M beam profiler from Thorlabs. The sensor has 1360 x 1024 pixels, and each pixel is of size ≈ 6.45$\mu$m x 6.45$\mu$m. This CCD is used along with Neutral Density (ND) filters of 20 dB or 40 dB based on our requirement (here, these dB values represent the attenuation of these filters).

Initially, the input beam gets split equally by BS1. The retroreflectors of both arms are kept perpendicular to each other. RR1 flips the beam profile from $r(x,y)$ to $r(x,-y)$ and RR2 will flip it from $r(x,y)$ to $r(-x,y)$. These two flipped beams are made to overlap using the mirror arrangements, shown in Fig.1. After BS2 the combined beams are incident on the CCD camera. Then in the image plane of CCD, each pixel will correspond to the spatial correlation between the points $r(x,y)$ of the original beam and $r(-x,-y)$ after being reflected from both RR1 and RR2. So each point in the image plane correspond to an effective spatial separation $(\delta x, \delta y) = (2|x|, 2|y|)$. Now, upon overlap, an interference fringe pattern is observed as shown in Fig. 2(a) and 2(b). The fringe pattern also depends on the tilt introduced in M2. This tilt of the M2 mirror is controlled in our case by moving it by a distance $\Delta$ and then rotating the mirror handle (lying along x axis) about y axis like in Fig. 1(b) to make them interfere on the screen of the CCD. This can also be done by rotating the mirror along any other axes. The larger is the $\Delta$, smaller is the fringe width. Therefore, this control can help us to measure coherence function of light sources/spots having smaller spatial patterns. One can always choose to measure fringe patterns which are at least 2 or 3 pixels wide on the CCD to prevent random fluctuations and pixel blurring in the measurement during the 2D mapping of $|g^{(1)}(\delta x, \delta y)|$.

To measure the degree of spatial coherence, first we vary temporal difference of the interfering light spots by moving the RR2 longitudinally around $\tau = 0$ to eliminate any residual temporal contributions from the spatial coherence measurements using 'temporal filtering'[24]. This is done using Piezoelectric Actuator PAS005 procured from Thorlabs and then we record the observed interference patterns. The point $\tau = 0$ is fixed by going to a less bias current value of the laser diode with negligible coherence. Here the interference is observed over only small temporal differences. Whenever clear fringes are observed, it is taken as $\tau = 0$ approximately. The Piezo is controlled using MDT693B 3-Channel open-loop Piezo controller from Thorlabs. The voltage applied is varied from 0.00 to 6.00 V in steps of 0.15 V resulting in a total movement of ≈16 $\mu$m in Piezo giving a temporal delay of ≈108 fs in total around $\tau$

= 0. This 'temporal filtering' is known[24] to eliminate the mixing of unwanted temporal coherence in spatial coherence measurement. The intensity of both beams were also recorded separately by blocking the other to find the values of $I_1$ and $I_2$. From these recorded interference patterns, once the intensity from the same pixel of the CCD was plotted as a function $\tau$, we get $I(\tau)$ vs $\tau$ for a particular ($\delta x, \delta y$) as shown in Fig. 2(d). Using equation 2, we then calculate the degree of spatial coherence $|g^{(1)}(\delta x, \delta y)|$ for each pixel corresponding to each ($\delta x, \delta y$) at $\tau = 0$ condition. Repeating it over all the pixels, we end up with the 2D map of $|g^{(1)}(\delta x, \delta y)|$. The same measurement is taken five times and the average of it is taken as the final value. The scattered background light outside the beam diameter can affect the Fourier transform in case the beam is partially coherent, like in the case of 20 mA bias current. So, the noise outside is removed by keeping only the values in the beam area shown in Fig. 2(c) and making all other values be 0. This also helps to avoid any lens vignetting issues around the peripheral regions. The coherence function is calculated for various bias currents and plotted in Fig. 3. In this experiment, we assume the beam centre to be $\delta x = 0$, $\delta y = 0$. Now taking the 2D Fourier transform of this $|g^{(1)}(\delta x, \delta y)|$, while keeping the point ($\delta x = 0$, $\delta y = 0$) as the centre, we then acquire the 2D k-space map of the emitted light. The calculated k-space for varying bias current is shown in Fig. 3. Under Fast Fourier Transform (FFT) any function $f(cx) \xrightarrow{FFT} \frac{1}{|c|} F\left(\frac{k}{c}\right)$, c is a constant which is 2 as argued above. So we multiply the k-space coordinates with |c|=2 to determine $F(k)$ vs k plots. So only the amplitude of the real part of the obtained Fourier transform F(k) is plotted following Eq. 5.

## III. RESULTS AND DISCUSSIONS

From Fig. 3 we can see that on varying bias current, the coherence of the beam increases in magnitude monotonically. At 20 mA and 22.5 mA the laser diode is in LED mode and we can see that the measured degree of coherence is very low. At 25 mA, which is the threshold bias current for lasing, there is an onset of coherence, and at 27.5 mA and 30 mA. Here, the coherence function is highly enhanced, which is due to increase in stimulated emission or lasing. The 2D k-space plots, after being normalised as $[M]' = \frac{[M]-\min(M)}{\max(M)-\min(M)}$ where the matrix [M] is the FT of 2D coherence function, for varying bias current are also shown in Fig. 3 respectively. We can see that the occupations of high k-values are decreased and of lower k-values are increased as the bias current increases. As a result of stimulated emission after lasing, light emission from the diode becomes more directional along with associated changes in the k-space. While varying bias current, the intensity of the beam also increases, so the exposure time and gain of the CCD are modified while using different ND filters. However, these CCD parameters are maintained at a constant level during the measurements for the same bias current. So, the calculated $|g^{(1)}(\delta x, \delta y)|$ do not depend on exposure time and gain during comparisons shown in Figs. 3(a)-3(e), because these changes cancel out in Eq. 2. In our experiment, we had a smaller number of points within k-space. To increase the sampling frequency in k-space ($\Delta_f$), we increased the maximum sample size of the coherence function ($X_{max}$), by adding additional empty columns to both sides of obtained coherence function and also this won't affect the obtained coherence function.

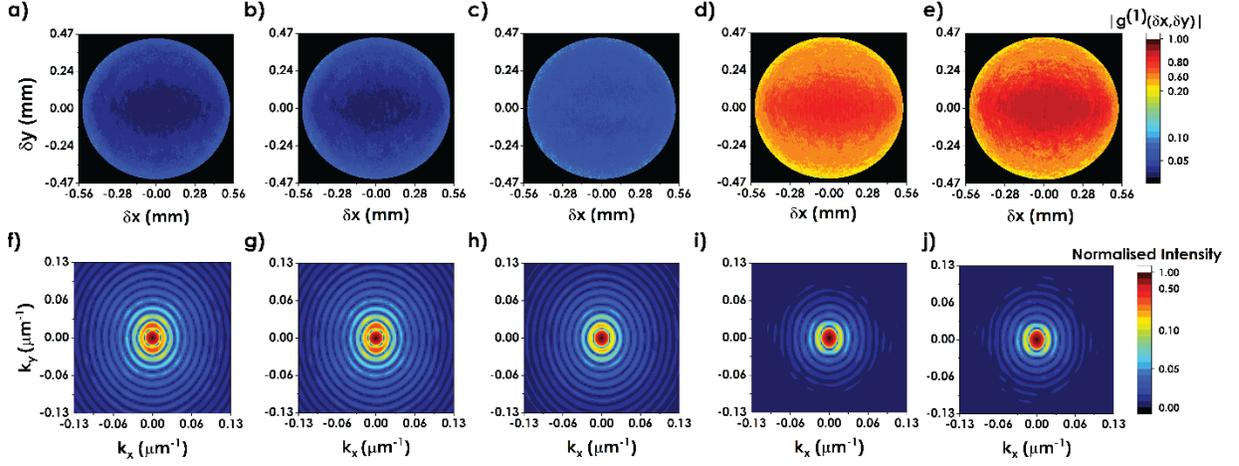

**FIG. 3.** (a)-(e) show the degree of coherence function measured for 20 mA, 22.5 mA, 25 mA, 27.5 mA and 30 mA respectively. (f)-(j) show the calculated 2D in-plane momentum space for the above mentioned bias currents respectively using Eq. 5.

## A. VALIDATING THE OBTAINED MOMENTUM SPACE DISTRIBUTION FROM COHERENCE METHOD

To show what we estimated as 2D momentum space map is indeed the 2D momentum space distribution of the emitted light, we tried to measure the k-space of a double slit with slit separation 0.25 mm and slit width 0.2 mm. We then compare it with the usual Back Focal Plane Imaging of the same and also with a numerical simulation of the k-space of the double slit.

### 1. Theory

*a. Back Focal Plane Imaging*

Generally, the back focal plane imaging is done using a lens to do the Fourier transform as in Fig. 4(a). Here the f is the focal length of the lens L2. From this figure we can tell that

$$\frac{x_f}{f} = \frac{-x}{S}$$
$$x_f = \frac{-x \times f}{S}$$

(6)

where $\frac{-x}{S} = \alpha$ is the directional cosine[40] and given as $\alpha = -u \times \lambda$, $\lambda$ is the wavelength of given light and u is the spatial frequency. So, $x_f = -u \times \lambda \times f$.

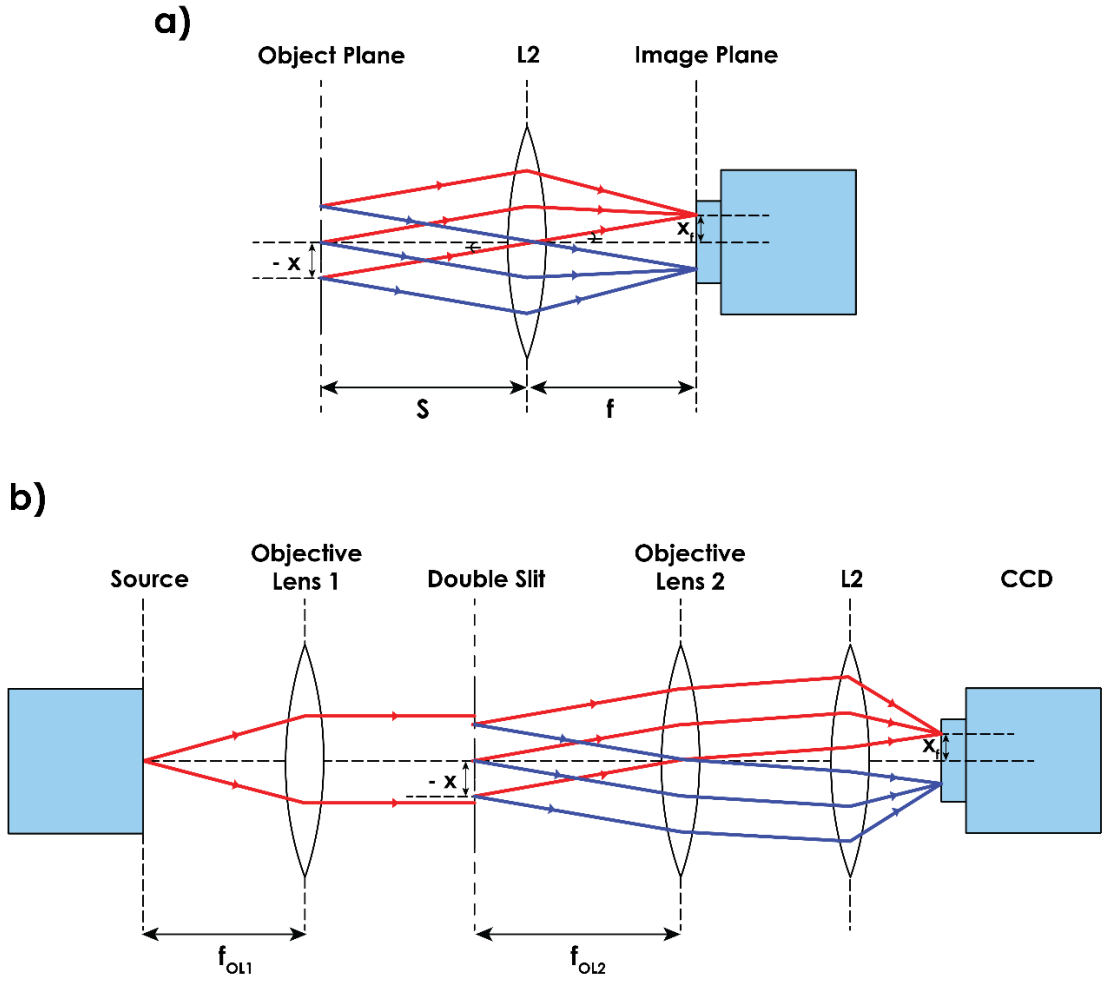

**FIG. 4.** (a) shows the standard setup used for BFP imaging. (b) shows our arrangement for obtaining BFP. The focal length of two objective lens are **f_OL1** and **f_OL2** respectively.

$$u = -\frac{x_f}{\lambda \times f}$$

(7)

Here the negative sign is used because the lens inverts the image. Therefore, the angular spatial frequency is calculated as

$$k_{bfp} = (-1)\frac{x_f (2\pi)}{\lambda \times f} \qquad (8)$$

Since we are trying to find k-space of double slit which can give rise to interference pattern, we used another objective lens whose focal plane has the double slit as shown in Fig. 4(b). As a result, its magnification factor is also taken into consideration in the calculation of the Fourier space coordinates. The double slit is an extended source so the objective lens can't collimate it fully and we end up with a diverging beam. To rectify this, a correction factor is taken into the Fourier coordinates as the ratio of the observed slit width at the plane of L2 to the expected slit width for an objective lens of known magnification. However, the rectangular slit function along the y-direction is much larger compared to its dimension along x. Therefore, in case of double slits, we are mostly concerned about the spread of momentum/wave-vector in the x-direction. So this correction factor has not been used for y-direction to determine the k-space map.

*b. Fourier Transform based coherence method*

In our optical coherence method, the Fourier coordinates are determined by the program used for the FFT. To check the obtained k-space from BFP and coherence method, we also developed a simulation for the k-space. The real space information is given as a matrix, say $A_{p,q}$ run from 0 to N-1 and 0 to M-1 respectively where N/M is the maximum no. of data points taken in the x or y Cartesian coordinate respectively. The sampling interval (D) is either determined experimentally from pixel width of the CCD. However, this interval is set to a particular value during simulation, so that each index ($p$) would correspond to the real space coordinate $p \times D$ and the entire range is $X_{max} = N \times D$. The 2D FFT performed in python is given as,

$$F_{u,v} = \sum_{p=0}^{N-1}\sum_{q=0}^{M-1} A_{p,q} e^{-i2\pi p.u} e^{-i2\pi q.v} \qquad (9)$$

where u and v are indices of Fourier space matrix. The sampling frequency ($\Delta_f$) is given as $\Delta_f = \frac{1}{D} = \frac{N}{X_{max}}$. The Fourier coordinates or resulting spatial frequencies (U) are given as, $U = \frac{u}{N}\Delta_f = \frac{u}{X_{max}}$ and the corresponding angular spatial frequency ($k_x$) is given as $k_x = \frac{u \times 2\pi}{X_{max}}$ and similarly we can write $V = \frac{v}{Y_{max}}$ thus $k_y = \frac{v \times 2\pi}{Y_{max}}$. Also if we have used an objective lens of 10X Magnification, then the above formula becomes, $k_x = \frac{u \times 2\pi}{(X_{max})/10}$ and similarly for $k_y$.

In this method, the maximum frequency detectable is $\frac{\Delta_f}{2}$. In our experiment $X_{max}$ = pixel size × (Total no. of pixels taken along x) × 2 × $1/Magnification$. As mentioned in the last sentence of Sec. II, a factor of 2 is also multiplied with the k-space coordinates while using the coherence method to determine F(k) vs k plots.

For double slit, the simulated k-space is the square of the real part of the Fourier transform[31] of its slit function, since the k-space is nothing but the diffraction pattern of the slit. The coordinates of the k-space here are decided exactly the same as mentioned above in this section. In the simulation, we created a matrix with each successive column or row corresponding to D = 10 μm for the slit function and its Fourier transform was obtained.

## 2. Using Double Objective Lens for the Optical Coherence Method

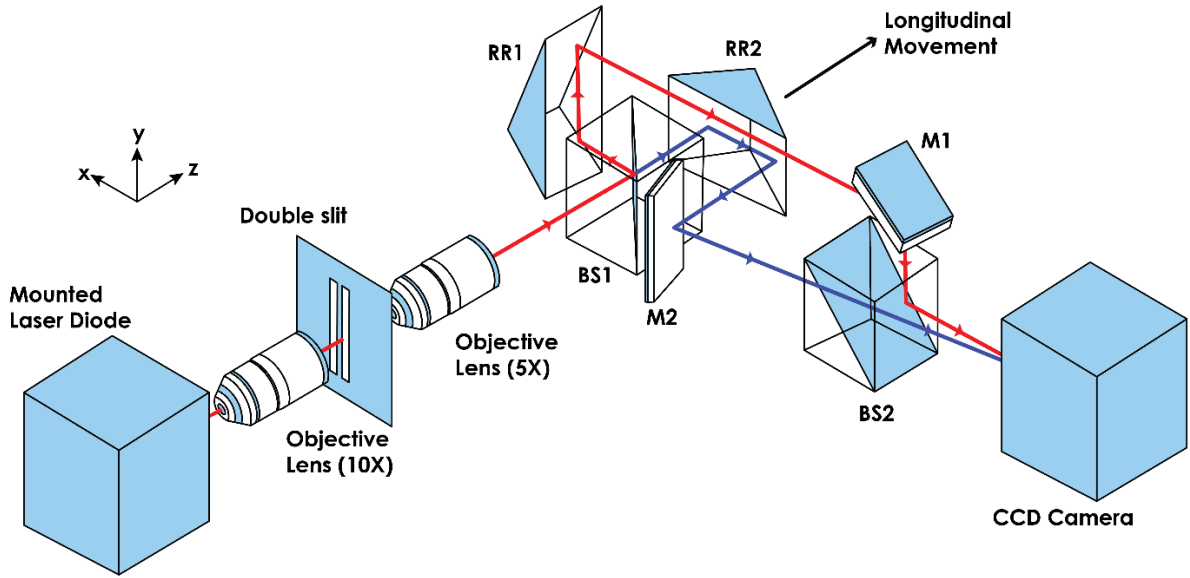

**FIG. 5.** This image shows the modification made to the setup in case of measuring coherence function for double slit.

The interference pattern from the double slits diverges with varying distance from the plane of slits and certainly not collimated to be used directly in the Michelson interferometer for measuring spatial coherence function $|g^{(1)}(\delta x, \delta y)|$ in a direction transverse to the light propagation direction z. So, if we measure the coherence function of this light coming out from the slit, it won't be the desired $|g^{(1)}(\delta x, \delta y)|$ of the double slit. To overcome this problem, another collimating lens is added after the double slit as shown in Fig. 5. The double slit is placed at the focal plane of this second objective lens so that the image of the double slit is collimated and then the beam is sent for measurement of the $|g^{(1)}(\delta x, \delta y)|$. However, the double slit is an extended source and the beam has some amount of divergence. As a result, a similar correction factor is also used with Fourier coordinates as the ratio of the observed slit width at the CCD to the expected slit width using an objective lens of known magnification. In our experimental arrangement, this was measured to be 0.46 (approx.). As mentioned above, this correction factor along y is not taken into consideration here. Also only the magnification of the second lens is taken into account since, in this case, the source itself is the double slit. The obtained k-space intensity from the $|g^{(1)}(\delta x, \delta y)|$ with the background noises removed, is plotted in Fig. 6a along with the simulation and the back focal plane for comparison. In that plot, the Fourier coordinates for optical coherence method have an additional factor of 2 which was explained above in Sec. II. In our experiment, we had a smaller number of points within k-space. To increase the sampling frequency in k-space ($\Delta_f$), we increased the maximum sample size of the coherence function ($X_{max}$), by adding additional empty columns to both sides of obtained coherence function and also this won't affect the obtained coherence function.

## 3. Comparisons

The Figs. 6a-6c show the plot of the 2D k-space obtained from by our spatial coherence method (D = 0.59 μm, $X_{max} = 2.6\ mm$ ) as well as the simulation of the double slit (D = 10 μm, $X_{max} = 10\ mm$), measured using standard BFP method above the lasing threshold of the diode. While comparing them, we had to take into account the divergence of the light beam which arises from the extended nature of the source. In Fig. 6d, we compare the $k_x$ profile of the coherence method at $k_y = 0$ with the simulation and also with that acquired directly using of BFP method after taking the beam divergence into consideration. In all these three methods, the correction factor along $k_y$ is not taken into account since our focus was mainly along $k_x$. As a result, it leads to somewhat different spreads along $k_y$ upon comparison. We see narrower spread along $k_y$ in case of coherence method because the measured coherence map $|g^{(1)}(\delta x, \delta y)|$ fills up the entire sample space of our CCD due to the beam divergence. In case of simulation, we also assumed the slits to be extended along y and as a result the spread along $k_y$ is not much. Whereas in the BFP imaging, slit image obtained after second objective lens is completely collected by the lens L2 and we see somewhat larger spreads along $k_y$. The optical elements used in our setup are simple components compared to previously proposed setups and can be aligned much more easily. The surfaces of retroreflectors used in our setup has 'no discernible' obliquity other than the right angle maintained between two internal reflecting surfaces. So, polarisation dependent phase shifts in the light beam are prevented. Also, there is no spherical surfaces used in our system which can introduce additional asymmetric/non uniform path lengths or phase differences. Moreover, in our setup, we have a control over the additional phase which we introduce to the beam. This can help whenever we are imaging smaller sized micro/nano structures. We ensured that considerable number of fringes (say ~10 as shown in Fig. 2a) were used to determine the coherence function $|g^{(1)}(\delta x, \delta y)|$.

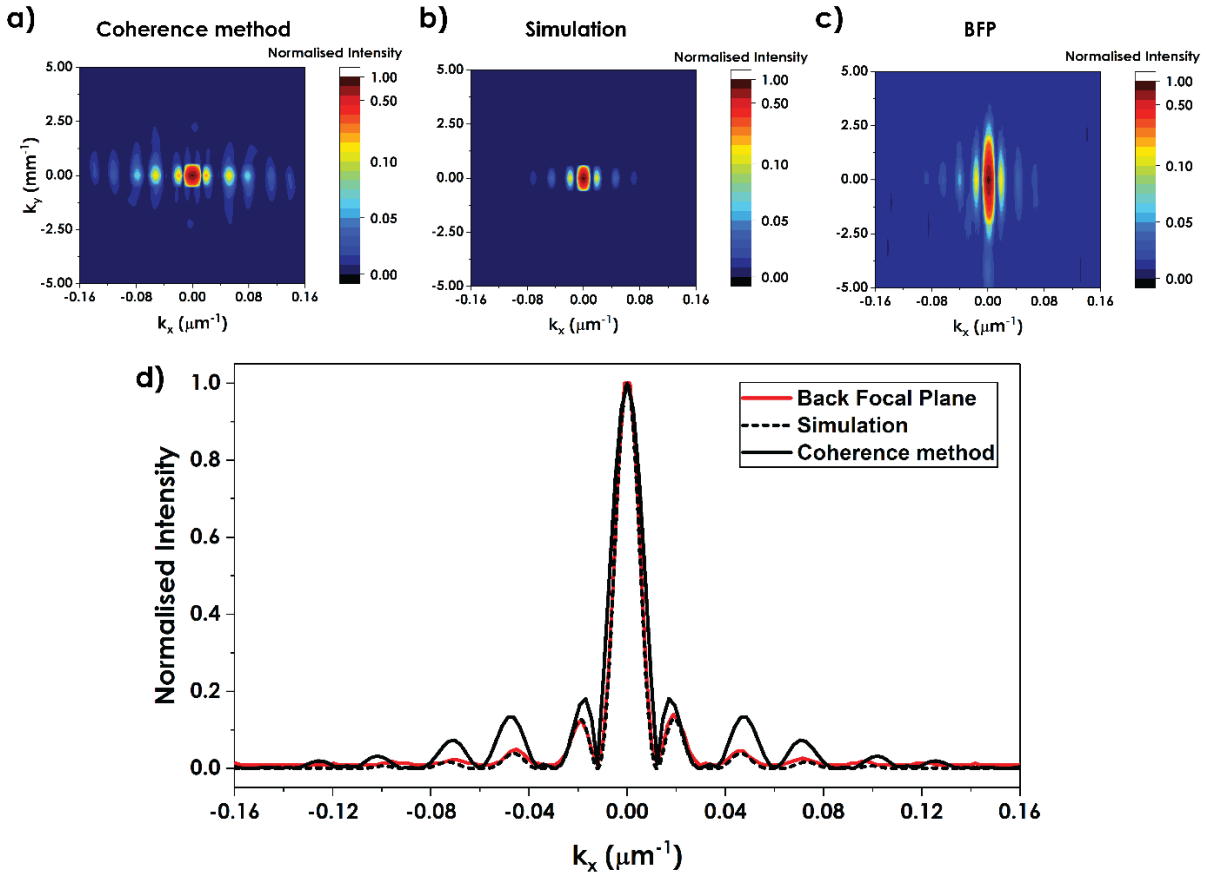

**FIG. 6.** (a) This shows the momentum space obtained as the amplitude of the real part of Eq. 5 using our coherence method as shown in Fig. 5. (b) It shows the simulation as square of the real part of the Fourier transform of the slit function. (c) Shows the BFP image measured using the arrangements shown in Fig. 4b. (d) This compares only $k_x$ profiles of the above plots. While plotting the momentum coordinates for coherence method, we have considered a multiplying factor of 2 as explained in the text.

We can also change the direction in which the fringes are formed by adjusting either of M1 or M2, so that it can be optimised at ease for any further measurements of interests. The k-space calculated from the coherence function is shown to be the actual momentum space by comparing the results from measuring k-space of a double slit and the simulation of the k-space of a double slit. We also compared 2D k-space measured using this coherence method with the one obtained from the usual Back Focal Plane Imaging method using a common simulation.

## IV. CONCLUSION

We have demonstrated the use of a modified Michelson Interferometer to determine the 2D momentum space distribution of light. In this, first we collimate the light and then measure its degree of 2D transverse spatial coherence over a selected area using that modified optical interferometry setup which can be kept outside the low temperature cryostat. This areal mapping of the first order, spatial coherence function is obtained by finding the visibility factor from observed interference fringes using the required temporal filtering method[24] for each spatial coordinate around $\tau = 0$. Finally, we calculate the 2D Fourier transform of this 2D map of first order correlation function which allowed us to obtain the in-plane momentum space distribution of light. As mentioned earlier, in this method if we want to increase the resolution of the 2D momentum space map, then the range over which the coherence function $|g^{(1)}(\delta x, \delta y)|$ is measured has to be increased.

Our experimental setup has an advantage in its capacity to control the fringe width by simply tilting only one mirror (say M2), thus enabling us to study even smaller emission pattern over a 2D area. It doesn't have problems associated with the use of CER [35-37], dove prisms[38] and also with those studies where only 2D coherence function was studied but not the momentum space. There were momentum space studies[27] with spatial coherence measurement but only in 1D. However, in this paper we measured the full 2D map of coherence function over a selected area of the light beam and then used that directly to determine the 2D momentum space, which was not explored earlier in this particular way. We have used this setup to measure the 2D coherence function as well as the 2D momentum space distribution of light from a laser diode as a function of its bias voltage and observed the evolution of the momentum space during the onset of lasing. Moreover, this method is certainly simpler and advantageous in cases where the sample under study is kept at a low temperature, but k-space distributions of light emission can still be measured using optical instrumentations placed totally outside the low temperature cryostat. Therefore, this experimental design will certainly be helpful to probe Bose-Einstein Condensate of excitons, polaritons, photons by detecting the narrowing of momentum space of emitted light as an evidence of the underlying long-range spatial coherence or order.

**SUPPLEMENTARY MATERIAL**

See the Supplementary Material for the python code that was used for simulation of momentum space of the double slit pattern The matrix of size $1000 \times 1000$ is taken over the area of 10 mm $\times$ 10 mm with each row or column representing 0.01 mm. The obtained Fourier transform is then normalised for further comparison. We have also provided the python code for the calculation of $\left|g^{(1)}(\delta x, \delta y)\right|$ and its FFT to determine the 2D k-space.

**ACKNOWLEDGMENTS**

SD acknowledges the Science and Engineering Board (SERB) of Department of Science and Technology (DST), India (Grants # DIA/2018/000029, CRG/2019/000412) for supports. We thank Prof. K. L. Narasimhan for useful discussions.

**AUTHOR DECLARATIONS**

**Conflict of Interest**

Authors declare no conflicts on interests.

**Author Contributions**

S. V. U. Vedhanth (SVUV) and Shouvik Datta (SD) had planned and designed the experiments. SVUV had taken the data and analyzed those using simulations. Both SVUV and SD wrote the manuscript together.

**DATA AVAILABILITY**

All data and materials used in the analyses will be available for support the findings of this study and/or for purposes of reproducing and/or extending these analyses from the corresponding author upon reasonable requests.

**REFERENCES**

[1] A. Imamoglu, R.J. Ram, S. Pau, and Y. Yamamoto, Phys. Rev. A **53**, 4250 (1996).

[2] R. Matsuzaki, H. Soma, K. Fukuoka, K. Kodama, A. Asahara, T. Suemoto, Y. Adachi, and T. Uchino, Phys. Rev. B **96**(12), 125306 (2017).

[3] Z.K. Tang, M. Kawasaki, A. Ohtomo, H. Koinuma, and Y. Segawa, J. Cryst. Growth **287**, 169 (2006).

[4] Y.-Y. Lai, Y.-H. Chou, Y.-P. Lan, T.-C. Lu, S.-C. Wang, and Y. Yamamoto, Sci. Rep. **6**, 20581 (2016).

[5] T. Shih, E. Mazur, J.-P. Richters, J. Gutowski, and T. Voss, J. Appl. Phys **109**, 043504 (2011).

[6] G.C. La Rocca, Nat. Photonics. **4**, 343 (2010).

[7] P. Bhattacharya, T. Frost, S. Deshpande, M.Z. Baten, A. Hazari, and A. Das, Phys. Rev. Lett. **112**, 236802 (2014).

[8] C. Schneider, A. Rahimi-Iman, N.Y. Kim, J. Fischer, I.G. Savenko, M. Amthor, M. Lermer, A. Wolf, L. Worschech, V.D. Kulakovskii, I.A. Shelykh, M. Kamp, S. Reitzenstein, A. Forchel, Y. Yamamoto, and S. Höfling, Nature **497**, 348 (2013).

[9] T. Damm, D. Dung, F. Vewinger, M. Weitz, and J. Schmitt, Nat. Commun. **8**, 158 (2017).

[10] J. Klaers, J. Schmitt, F. Vewinger, and M. Weitz, Nature **468**, 545 (2010).

[11] J.P. Eisenstein, and A.H. MacDonald, Nature **432**, 691 (2004).

[12] L. V. Butov, C.W. Lai, A.L. Ivanov, A.C. Gossard, and D.S. Chemla, Nature **417**, 47 (2002).

[13] T. Byrnes, N.Y. Kim, and Y. Yamamoto, Nat. Phys. **10**, 803 (2014).

[14] D.W. Snoke, and J. Keeling, Phys. Today **70**, (2017).

[15] F. Scafirimuto, D. Urbonas, U. Scherf, R.F. Mahrt, and T. Stöferle, ACS Photonics **5**, (2018).

[16] T. Byrnes, N.Y. Kim, and Y. Yamamoto, Nat. Phys. **10**, (2014).

[17] V.P. Ryabukho, A.L. Kal'yanov, D. V. Lyakin, and V. V. Lychagov, Opt. Spectrosc. **108**, 979 (2010).

[18] D. Morrill, D. Li, and D. Pacifici, Nat. Photonics **10**, 681 (2016).

[19] T.E.C. Magalhães, J.M. Rebordão, and A. Cabral, Optik **226**, 166034 (2021).

[20] H. Deng, G.S. Solomon, R. Hey, K.H. Ploog, and Y. Yamamoto, Phys. Rev. Lett. **99**, 126403 (2007).

[21] F. Manni, K.G. Lagoudakis, B. Pietka, L. Fontanesi, M. Wouters, V. Savona, R. André, and B. Deveaud-Plédran, Phys. Rev. Lett. **106**, 176401 (2011).

[22] K.S. Daskalakis, S.A. Maier, R. Murray, and S. Kéna-Cohen, Nat. Mater. **13**, 271 (2014).

[23] A. Bhattacharjee, S. Aarav, and A.K. Jha, Appl. Phys. Lett. **113**, 051102 (2018).

[24] M. K. Singh, and S. Datta, Rev. Sci. Instrum. **92**, 105109 (2021).

[25] J. Marelic, L.F. Zajiczek, H.J. Hesten, K.H. Leung, E. Y X Ong, F. Mintert, and R.A. Nyman, New. J. Phys. **18**, 103012 (2016).

[26] A. Efimov, Opt. Express. **22**, 15577 (2014).

[27] S. Yang, A.T. Hammack, M.M. Fogler, L. V. Butov, and A.C. Gossard, Phys. Rev. Lett **97**, 187402 (2006).

[28] A.H. Schokker, and A.F. Koenderink, Phys. Rev. B. **90**, (2014).

[29] A.B. Vasista, D.K. Sharma, and G.V.P. Kumar, in *Digital Encyclopedia of Applied Physics* (Wiley-VCH Verlag GmbH & Co. KGaA, Weinheim, Germany, 2019), pp. 1–14.

[30] Y. Zhang, M. Zhao, J. Wang, W. Liu, B. Wang, S. Hu, G. Lu, A. Chen, J. Cui, W. Zhang, C.W. Hsu, X. Liu, L. Shi, H. Yin, and J. Zi, Sci. Bull. **66**, 824 (2021).

[31] Eugene Hecht, *Optics*, 5th ed. (Pearson, 2020), pp. 588-603 on optical coherence and pp. 483-487 on diffraction pattern as square of the real part of Fourier Transform of the slit function.

[32] Mark Fox, *Quantum Optics: An Introduction* (OUP Oxford, 2006), pp. 15-19.

[33] Rodney Loudon, *The Quantum Theory of Light*, 3rd ed. (Oxford Science Publications, 2000), Ch. 3, pp. 94-103.

[34] T.B. Hoang, G.M. Akselrod, A. Yang, T.W. Odom, and M.H. Mikkelsen, Nano. Lett. **17**, 6690 (2017).

[35] O. Hofherr, C. Wachten, C. Müller, and H. Reinecke, in *SPIE 8466, Instrumentation, Metrology, and Standards for Nanomanufacturing, Optics, and Semiconductors VI*, edited by S. Han, T. Yoshizawa, and S. Zhang (2014), p. 92760V.

[36] O. Hofherr, C. Wachten, C. Müller, and H. Reinecke, in *SPIE 8466, Instrumentation, Metrology, and Standards for Nanomanufacturing, Optics, and Semiconductors VI*, edited by M.T. Postek, V.A. Coleman, and N.G. Orji (2012), p. 84660J.

[37] E.R. Peck, J. Opt. Soc. Am. **52**, 253 (1962).

[38] I. Moreno, G. Paez, and M. Strojnik, Appl. Opt. **42**, 4514 (2003).

[39] F.E. Peña-Arellano, and C.C. Speake, Appl. Opt. **50**, 981 (2011).

[40] Miles V. Klein, *Optics*, 2nd ed. (Wiley, 1986), pp. 444-468.

- Supplementary Material -

# Measurement of 2D Momentum Space from 2D Spatial Coherence Measured using a Modified Michelson Interferometer

S. V. U. Vedhanth and Shouvik Datta [a]

*Department of Physics, Indian Institute of Science Education and Research, Pune 411008, Maharashtra, India*

Authors to whom correspondence should be addressed:

a) [shouvik@iiserpune.ac.in](mailto:shouvik@iiserpune.ac.in)

## 1) SIMULATION OF DOUBLE SLIT

```
# This is a python code for obtaining k-space of double slit from the Fourier
transform of its slit function

# Libraries needed

import numpy as np
import matplotlib.pyplot as plt

# Creating the double slit pattern by taking distance between each pixel to be
0.01 mm. So 0.2 mm (slit width) corresponds to 20 pixels and 0.25 mm (slit
separation) corresponds to 25 pixels

slit = np.empty((1000,1000))
for i in range(0,1000):
    for j in range(0,1000):
        if ((476<j<498) and (1<i<1000)) or ((502<j<524) and (1<i<1000)):
            slit[i,j] = 1
        else:
            slit[i,j] = 0

# Taking its Fourier Transform

diff = np.fft.fft2(slit)
diff = np.fft.fftshift(diff)
real = np.real(diff)
diff = (real)**2

# Normalising the k-space

Imin = np.min(diff)
Imax = np.max(diff)
normalized_matrix = (diff - Imin) / (Imax - Imin)
```

# Plotting

```
plt.subplot(121)
plt.imshow(slit)
plt.subplot(122)
plt.imshow(normalized_matrix)
plt.show()
```

## 2) CODE FOR CALCULATING $|g^{(1)}(\delta x, \delta y)|$ AND K-SPACE FROM THE OBSERVED INTERFERENCE FRINGES

# Importing required Libraries

```
import pandas as pd
import glob
import numpy as np
import math
import xlsxwriter
import matplotlib.pyplot as plt
import statistics
from statistics import mean
```

# To check for errors

```
no_of_mistakes = []
file_check = []
int_check = []
```

# Defining Visibility function

```
def Vis(x,A,B,i,l):
    y = ((max(x[0:30,i])-min(x[0:30,i]))/(max(x[0:30,i])+min(x[0:30,i])))*((A[l,i]+B[l,i])/(2*np.sqrt(A[l,i]*B[l,i])))
    return y
```

# Defining the path the files are in. Here the program is written for case where the main folder has all 5 attemts' raw data in separate folders and each of those folders have a subfolder for the intensities $I_1$ and $I_2$

```
path = r'<path>' # use your path
all_files = glob.glob(path + "/*.csv")
file_check.append(len(all_files))
li = []
```

# Reading all files

```
for filename in all_files:
    df = pd.read_csv(filename, delimiter=';')
    li.append(np.array(df))
```

# Reading I1 and I2

```
path2 = r'<path>\intensities' # use your path
```

```python
all_files2 = glob.glob(path2 + "/*.csv")
int_check.append(len(all_files2))
li3 = []

for filename in all_files2:
    df2 = pd.read_csv(filename, delimiter=';')
    li3.append(np.array(df2))

print(len(li3))
I1=li3[0]
I2=li3[1]

# Taking values of each pixel from different τ and calculation of $|g^{(1)}|$

li2=[]
for j in range (0,1023):
    m=0
    matrix = np.empty((41,1360))
    for x in li:
        for i in range (0,1359):
            matrix[m,i] = x[j,i]
        m+=1
    li2.append(matrix)
    print(j)

print(len(li2))

g1 = np.empty((1024,1360))
l=0
k=0
for x in li2:
    for i in range(0,1359):
        g1[l,i] = Vis(x,I1,I2,i,l)
        if g1[l,i]>=1:
            k+=1
    l+=1
    print('done1')

no_of_mistakes.append(k)
print('done 1')

############################################################################
# Repeating the above for few more times by using the following subroutine
successively.

path = r'<path>' # use your path
all_files = glob.glob(path + "/*.csv")
file_check.append(len(all_files))
li = []

for filename in all_files:
    df = pd.read_csv(filename, delimiter=';')
    li.append(np.array(df))

path2 = r'<path>\intensities' # use your path
```

```python
all_files2 = glob.glob(path2 + "/*.csv")
int_check.append(len(all_files2))
li3 = []

for filename in all_files2:
    df2 = pd.read_csv(filename, delimiter=';')
    li3.append(np.array(df2))

print(len(li3))
I1=li3[0]
I2=li3[1]

li2=[]
for j in range (0,1023):
    m=0
    matrix = np.empty((41,1360))
    for x in li:
        for i in range (0,1359):
            matrix[m,i] = x[j,i]
        m+=1
    li2.append(matrix)
    print(j)

print(len(li2))

g2 = np.empty((1024,1360))
l=0
k=0
for x in li2:
    for i in range(0,1359):
        g2[l,i] = Vis(x,I1,I2,i,l)
        if g2[l,i]>=1:
            k+=1
    l+=1
    print('done2')

no_of_mistakes.append(k)
print('done 2')

################################# AVERAGING #################################

# Averaging over all 5 attempts

gavg = np.empty((1024,1360))
gstd = np.empty((1024,1360))
m=0
for i in range(0,1023):
    for j in range(0,1359):
        glist = [g1[i,j],g2[i,j],g3[i,j],g4[i,j],g5[i,j]]
        gavg[i,j] = mean(glist)
        gstd[i,j] = statistics.stdev(glist)
        if gavg[i,j]>=1:
            m+=1
            gavg[i,j]=1
        print('done avg')
no_of_mistakes.append(m)
```

################## SELECTING THE SPATIAL RANGE AND FFT ###################

# Removing Lens Vignetting and associated noises around the Selected Area

```
g2 = np.empty((1024,4360))

for i in range(0,1359):
    for j in range(0,1023):
        if (((i-775)/415)**2 + ((j-515)/345)**2) <= 1:  # Define the area you
                                                        want the data
            g2[j,i] = g1.iloc[j,i]
            print('added')
        else:
            g2[j,i] = 0
            print('not added')
```

# Selecting a square area around where we selected the data for k-space calculation

```
g2 = g2[155:880,340:1210]
```

# Adding additional empty rows and columns to increase the resolution in k-space

```
g3 = np.empty((3000,3000))
for i in range(0,869):
    for j in range(0,724):
        g3[j+1137,i+1065] = g2[j,i]
        print('extending')
```

# Performing Fourier Transform

```
ft = np.fft.fft2(g3)
ft = np.fft.fftshift(ft)
ftt = np.sqrt((np.real(ft))**2)
```

# Normalising it

```
Imin = np.min(ftt)
Imax = np.max(ftt)
normalized_matrix = (ftt - Imin) / (Imax - Imin)
```

################# SAVING THE DATA IN EXCEL FILE AND PLOTTING #################

```
result = pd.DataFrame(normalized_matrix)
result.to_excel('./knorm.xlsx')

ga=pd.DataFrame(gavg)
ga.to_excel('./gavg.xlsx')

gs=pd.DataFrame(g3)
gs.to_excel('./gavg_selected.xlsx')

gst=pd.DataFrame(gstd)
gst.to_excel('./gstd.xlsx')
```

# Checking for errors, meaning if $|g^{(1)}|$ takes value other than range [0,1] or if any file is not read

```python
print('no of mistakes = ',no_of_mistakes)
print('no of Inter files = ',file_check)
print('no of I files = ',int_check)
```

# Plotting

```python
plt.subplot(131)
plt.imshow(gavg, cmap='viridis')
plt.subplot(132)
plt.imshow(g2, cmap='viridis')
plt.subplot(133)
plt.imshow(normalized_matrix, cmap='viridis')
plt.show()
```